\documentclass[prl,amsmath,amssymb,showpacs,twocolumn,superscriptaddress]{revtex4}
\usepackage{amsfonts}
\usepackage{amssymb}
\usepackage{latexsym}
\usepackage{graphicx}
\usepackage{epsfig}

\begin{document}

\title{A new class of three-body states}

\author{Nicolais Guevara}
\affiliation{Department of Physics, Kansas State University, Manhattan, Kansas, 66506, USA}

\author{Yujun Wang}
\affiliation{Department of Physics, Kansas State University, Manhattan, Kansas, 66506, USA}
\affiliation{JILA, University of Colorado, 440 UCB, Boulder, Colorado, 80309, USA}

\author{B. D. Esry}
\affiliation{Department of Physics, Kansas State University, Manhattan, Kansas, 66506, USA}

\begin{abstract} 

We calculate the three-body spectrum for identical bosons interacting via attractive $1/r^2$ potentials.
We have found an infinite number of three-body states even when the pair interactions are
too weak to support any two-body states.  These new states thus share this surprising scenario
with the Efimov effect, but are not themselves Efimov states.  Our effect occurs for both
identical bosons and identical fermions, and it persists in the presence of two-body bound
states. 

\end{abstract}

\maketitle

In its long history, quantum mechanics has offered a wide variety of counterintuitive phenomena.
One example that has received considerable interest lately 
is the Efimov effect from few-body
physics.  Predicted by Vitaly Efimov in 1970~\cite{Efimov1970,Efimov1971,Efimov1979,Efimov1973}, but not clearly observed experimentally 
until 2006 using ultracold atomic gases~\cite{Kraemer2006,Knoop2009,Ottenstein2008,Huckans2009,Zaccanti2009,Ferlaino2011}, 
the effect that has now taken his name refers 
to the emergence of an infinite number of three-body bound states 
when none of the two-body subsystems are bound.
Central to the effect is the requirement
that the two-body interactions are short-ranged, i.e. fall off faster than $1/r^2$ when $r$ is a
two-body interparticle distance, so that a two-body $s$-wave scattering length $a$ can be defined.
The Efimov effect occurs when $|a|$$\rightarrow$$\infty$.

Far from being an esoteric theoretical curiosity,
the Efimov effect has a profound impact on low-energy three-body collisions even when $a$
is finite~\cite{Esry1999,Nielsen1999,Braaten2006,Wang2010,Kraemer2006,Knoop2009,Ottenstein2008,Huckans2009,Zaccanti2009,Ferlaino2011}.
At low enough collision energies, the de~Broglie wavelength is large compared to 
a characteristic short-range length scale $r_0$ and thus 
insensitive to details on this scale.
Consequently, in this limit, three-body systems display universal behavior dependent only
on $a$~\cite{Braaten2006}.  
Therefore, the desire to obtain the universal behavior of
low-energy three-body observables motivates the study of weakly-bound three-body systems beyond
the fundamental, intrinsic interest such systems elicit.

Generally speaking, two-body potentials can be categorized as short- or long-ranged based on
whether they fall off faster or slower, respectively, than $1/r^2$ when $r$$\rightarrow$$\infty$.
Efimov's effect occurs in systems with short-range two-body interactions, and
we know that an infinity of three-body bound states can occur in systems with
long-range two-body interactions like the Coulomb interaction.
Thus, to complete our knowledge of three-body systems interacting
via local two-body potentials, we
must ask: What three-body spectrum results from attractive $1/r^2$ two-body interactions --- which lie exactly
at the boundary between short- and long-range?  

This question turns out to be quite rich.  For instance, depending on their strength, $1/r^2$
potentials can have no two-body bound states when they are subcritical or critical,
or they can have an infinity of two-body bound states when supercritical.
Clearly, one interesting
question is whether there is an infinite number of three-body bound states in the subcritical
case.  

In this Letter, we show that this scenario is indeed possible for three interacting, indistinguishable bosons in three
dimensions.  We can thus reproduce the counterintuitive result of Efimov's scenario, an infinity 
of three-body bound states in the absence of two-body bound states, but with {\em long-range} 
two-body potentials.
While our scenario bears a superficial resemblance to Efimov's, it is, however,
fundamentally different since $a$ --- a necessary ingredient for the Efimov effect --- is not even defined. 

In addition, we show that when two-body bound states {\em are} formed, either by a short-range interaction or 
by strongly attractive $1/r^2$ interactions, an infinite number of three-body bound and resonant states 
appear.
Moreover, we show that our
states also occur in a system of three identical fermions ---
a system for which no Efimov effect occurs.  On this basis, we argue that the four-body ``Efimov states'' 
recently identified in Ref~\cite{Castin2010} are actually 
our states.

Our study is based on 
numerical solutions~\cite{Suno2002} of the three-body Schr\"odinger equation in the adiabatic hyperspherical 
representation~\cite{Macek1968,Suno2002,NielsenReview}.
In this representation, the overall size of the system is measured by the hyperradius $R$, and the configuration 
of the system is represented by a set of five hyperangles collectively denoted $\Omega$.
Substituting the exact wave function 
in this representation,
\begin{equation}
\Psi(R,\Omega) = \sum_\nu F_\nu(R) \Phi_\nu(R;\Omega),
\label{ExactPsi}
\end{equation}
into the Schr\"odinger equation produces coupled hyperradial equations
\begin{equation}
\left(-\frac{\hbar^2}{2\mu}\frac{d^2}{d R^2}+W_{\nu\nu}\right)\!F_\nu+\sum_{\nu'\neq\nu}W_{\nu\nu'}F_{\nu'}=E F_\nu,
\label{HyperradEqn}
\end{equation}
where $\mu=m/\sqrt{3}$ for identical particles of mass $m$.  The adiabatic channel
functions $\Phi_\nu$ in Eq.~(\ref{ExactPsi}) are obtained from
\begin{equation}
H_{\rm ad} \Phi_{\nu}(R;\Omega) = U_{\nu}(R) \Phi_{\nu}(R;\Omega)
\label{AdEqn}
\end{equation}
in which the adiabatic Hamiltonian $H_{\rm ad}$ includes everything from the total
Hamiltonian except the hyperradial kinetic energy.

To determine the three-body spectrum, we focus on the behavior of $W_{\nu\nu}$ in Eq.~(\ref{HyperradEqn}).
We will restrict our discussion for three identical bosons to total orbital 
angular momentum $J$=0 and total parity $\Pi$=$+1$.
Moreover, it is sufficient to neglect the interchannel 
coupling $W_{\nu\nu'}$ since the single-channel approximation gives a strict upper bound to 
the three-body bound state energies~\cite{Starace1979}.
The $W_{\nu\nu}$ are defined as
\begin{equation}
W_{\nu \nu}(R)= U_{\nu}(R) + \frac{\hbar^2}{2 \mu}
 \left< \!\!\left< \frac{d\Phi_{\nu}}{dR} \biggl| \frac{d\Phi_{\nu}}{dR} \right> \!\! \right>.
\end{equation}
Here, the double-bracket signifies that the integration is carried out only over the hyperangles.

For the potential energy $V$ that appears in $H_{\rm ad}$,
we assume a pairwise sum of two-body potentials, $V$=$v(r_{12})$+$v(r_{13})$+$v(r_{23})$, where 
\begin{equation}
v(r_{ij})=-\frac{\alpha^2 +1/4}{m r_{ij}^2}\hbar^2
\label{Eq:Twobody}
\end{equation}
for interpartical distances $r_{ij}$.
For the subcritical case, --1/4$<$$\alpha^2$$\le$0~\cite{MorseFeshbach},
$v(r_{ij})$ does not support any two-body bound states 
even though it is attractive and long-ranged. For the supercritical case, $\alpha^2$$>$0,
the singularity at the origin leads to the ``fall-to-the-center'' 
problem, and a ground state is not defined~\cite{Thomas1935,Braaten2006}.
After some regularization, 
$v(r_{ij})$ supports an infinite number of two-body bound states with 
properties that generically follow a geometrical scaling.  For instance, the bound state
energy $E_n$ and mean radius $\langle r\rangle_n$ of the $n$-th state are, respectively,
\begin{equation}
 E_{n+1}/E_n = e^{-{2 \pi}/{ \alpha}}\quad {\rm and} \quad \langle r\rangle_{n+1} /\langle r\rangle_n  = e^{{\pi}/{ \alpha}}.
 \label{Eq:EfimovSpec}
\end{equation}

For the two-body potentials in Eq.~(\ref{Eq:Twobody}), it 
is well known that the hyperradial dependence can be exactly separated from the
hyperangles. 
In this case, $W_{\nu\nu}$=$U_\nu$ and takes the form 
\begin{equation}
W_{\nu\nu}=-\frac{\alpha_{\nu}^2+1/4}{2\mu R^2}\hbar^2,
\label{Eq:Universal}
\end{equation}
where $\alpha_\nu$ are universal constants dependent only on $\alpha$. 

For most any conceivable realization of an attractive $1/r^2$ two-body potential, however, there 
will be some short-range modification of Eq.~(\ref{Eq:Twobody}) that will remove the singularity.
To investigate the role of such a modification, 
we considered 
\begin{equation}
 v(r_{ij})= -\frac{\alpha^2 +1/4}{m [r_0^2 ~{\rm sech} ^2(r_{ij}/r_0)+r_{ij}^2]}\hbar^2.
 \label{Eq:TwobodyReg}
\end{equation}
The parameter $r_0$ sets the scale of the short-range regularization. 

\begin{figure}
\includegraphics[clip=true,width=0.95\columnwidth,height=0.4\columnwidth]{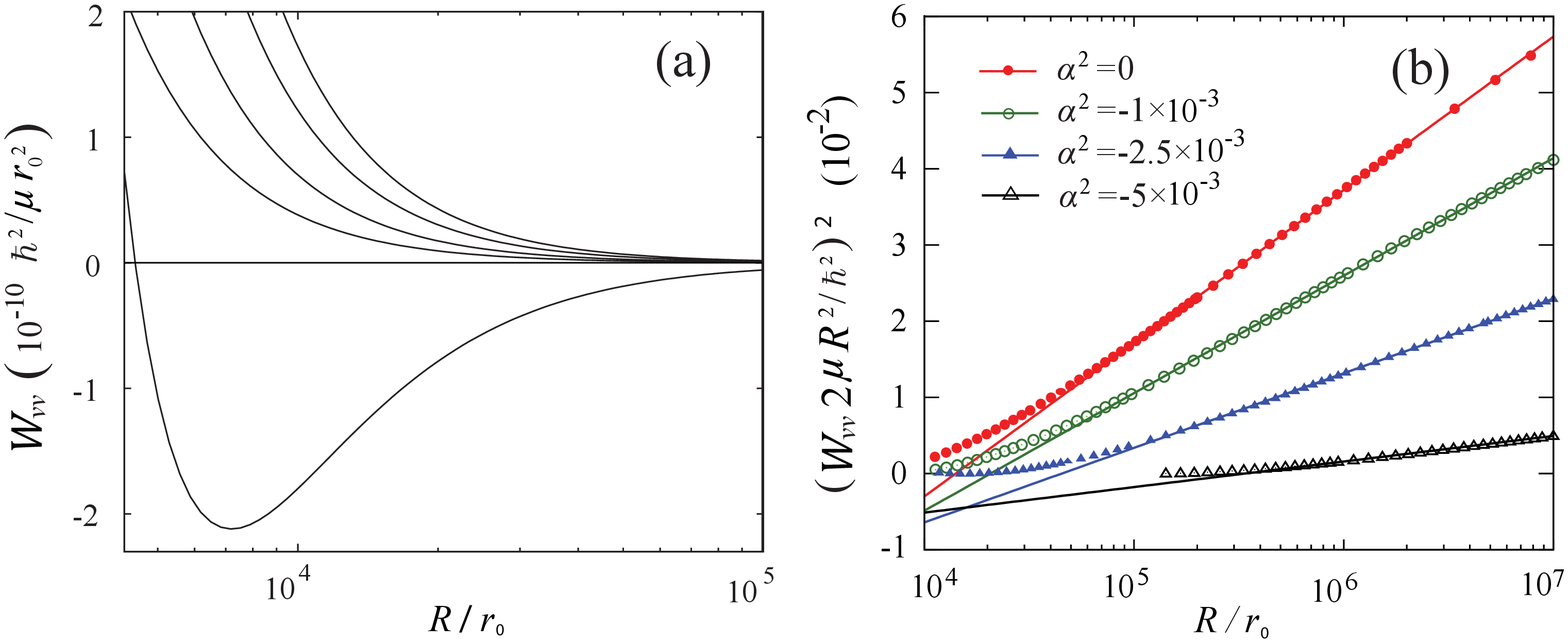}
\caption{(Color online.) (a) The three-body effective potentials $W_{\nu\nu}(R)$ for three identical bosons with $\alpha^2$=$0$. 
(b) The asymptotic behavior of $W_{00}(R)$ with the indicated values of $\alpha^2$.
The symbols are solutions of Eq.~(\ref{AdEqn}),
while the solid lines are fits based on Eq.~(\ref{Eq:LongRange}). 
\label{Fig:Potentials}}
\end{figure}

Surprisingly, for the subcritical regime --1/4$<$$\alpha^2$$\le$0, the regularized potential, Eq.~(\ref{Eq:TwobodyReg}), dramatically changes
the asymptotic behavior of $W_{00}$ from Eq.~(\ref{Eq:Universal}) to
\begin{equation}
 W_{00}(R)\rightarrow - \frac{\sqrt{\beta \ln(R/r_0)+\delta} }{2\mu R^2}\hbar^2.
 \label{Eq:LongRange}
\end{equation}
Note, however, that unlike Eq.~(\ref{Eq:Universal}), Eq.~(\ref{Eq:LongRange}) cannot 
currently be justified rigorously.  Instead, it is derived empirically from the numerical
results.  The quality of its asymptotic description is displayed 
in Fig.~\ref{Fig:Potentials}(b) where the
axes have been manipulated so that $W_{00}$ is a straight line if Eq.~(\ref{Eq:LongRange}) holds.  Moreover, 
by replacing ${\rm sech^2}(r_{ij}/r_0)$ in Eq.~(\ref{Eq:TwobodyReg}) by $\exp[-(r_{ij}/r_0)^2]$ 
and by unity, we have found that $W_{00}$ from Eq.~(\ref{Eq:LongRange}) is universal.
That is, $\beta$ is independent of the form of the short-range cutoff, but $\delta$ is not.
Equation~(\ref{Eq:LongRange}) thus displays 
another departure of our effect from Efimov's since
$r_0$=0 and $r_0$$\neq$0 have the same asymptotic $W_{\nu\nu}(R)$ in Efimov's case.

While the $R$ dependence of $W_{00}$ in Eq.~(\ref{Eq:LongRange}) is unusual, its consequence for
the number of three-body bound states is clear: $W_{00}$ falls off more slowly than $1/R^2$, making
it a long-range potential with an infinite number of three-body bound states.  
Figure~\ref{Fig:Potentials}(a), which shows the lowest several $W_{\nu\nu}$ for the most 
attractive case $\alpha^2$=0, also makes 
it clear that trimers produced by this effect are going
to be very weakly bound and very large --- the minimum in $W_{00}$ lies at roughly $R$=7000$r_0$.
As $\alpha^2$ decreases, $W_{00}$ becomes shallower, and the position of the minimum moves towards even larger $R$. 

Unfortunately, the $R$ dependence of $W_{00}$ in Eq.~(\ref{Eq:LongRange}) does eliminate the possibility of
finding an exact result like Eq.~(\ref{Eq:EfimovSpec}) for the three-body spectrum.
But, given the extremely slow variation of the numerator in Eq.~(\ref{Eq:LongRange}) and the fact that the 
hyperradial probability density is concentrated at the outer turning point,
we can treat $W_{00}$ as a pure $1/R^2$ potential with a variable coefficient based on the size of each state
to find an approximate expression for $E_n$. 
Replacing $R$ in the numerator of Eq.~(\ref{Eq:LongRange}) by $\langle R \rangle_n$ and 
using Eq.~(\ref{Eq:EfimovSpec}), we find for $n$$\gg$1
\begin{equation}
E_{n\!+\!1}/E_n=
  \exp\!\left(\!\!-\frac{2\pi}{[(\beta\ln\! \frac{\langle R \rangle_0}{r_0}\!-\!\frac{\beta}{2}\ln\! \frac{E_n}{E_0})^{1/2}\!-\! \frac{1}{4}]^{1/2}}\!\!\right)
\label{Eq:SubSpectrum}
\end{equation}
having neglected $\delta$.

\begin{figure}  
  \includegraphics[width=0.95\columnwidth,height=0.4\columnwidth]{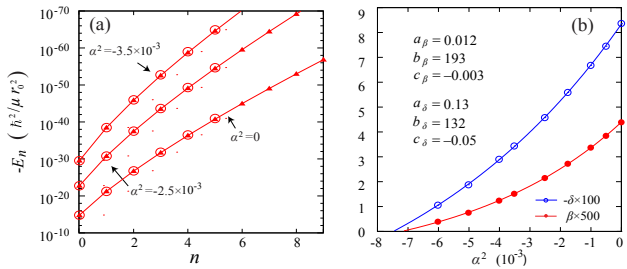}
  \caption{ (Color online.)
    (a) Comparison of the three-boson bound state energies calculated from Eq.~(\ref{HyperradEqn}) (circles)
    and from Eq.~(\ref{Eq:SubSpectrum}) (triangles).  The lines are added to guide the eye.
    (b) Fits of the potential parameters $\beta$ and $\delta$ from Eq.~(\ref{Eq:LongRange}): 
    symbols denote individual fits of Eq.~(\ref{Eq:LongRange}) to the numerical $W_{00}$; and 
    solid lines, fits of Eq.~(\ref{Eq:PPara}) to these values.
    \label{Fig:ParaFit}}
\end{figure}

Figure~\ref{Fig:ParaFit}(a) compares the three-body spectrum given by Eq.~(\ref{Eq:SubSpectrum})
with the energies calculated by solving 
Eq.~(\ref{HyperradEqn}) numerically using the potential given by Eq.~(\ref{Eq:LongRange}) cut off
by a hard wall at $R$=100$r_0$.
The agreement between the two is quite good, especially given that the energy scale covers 60
orders of magnitude, underscoring 
the value of
Eq.~(\ref{Eq:SubSpectrum}) since calculating many these states numerically is actually quite challenging.
The $\beta$ used in Fig.~\ref{Fig:ParaFit}(a) were determined by fitting 
the empirical expressions 
\begin{equation}
\beta=a_\beta\exp(b_\beta \alpha^2)+c_\beta,\;\;
-\delta=a_\delta\exp(b_\delta \alpha^2)+c_\delta.
\label{Eq:PPara}
\end{equation}
to the numerical data as shown in Fig.~\ref{Fig:ParaFit}(b). 

An infinity of three-body bound states thus exists at $\alpha^2$=0, and
Fig.~\ref{Fig:ParaFit}(b) shows that an infinity also exists for some negative $\alpha^2$.
Figure~\ref{Fig:ParaFit}(b) further shows that $\beta$ crosses zero at a critical value
$\alpha_c^2$ that can be determined from Eq.~(\ref{Eq:PPara}) 
to be $\alpha_c^2$$\approx$--0.0072,
However, due to the difficulty in solving Eq.~(\ref{AdEqn}) numerically to the large
distances required to identify the asymptotic behavior --- at least $R$=10$^9r_0$
near $\alpha_c^2$ --- this critical value should be treated as an approximate
upper bound.  The true value of $\alpha_c^2$ may be even more negative.
We have thus found an infinite number of 
three-body states for 
a range of $\alpha^2$ 
where no two-body states exist.  

\begin{figure}
  \includegraphics[width=0.95\columnwidth]{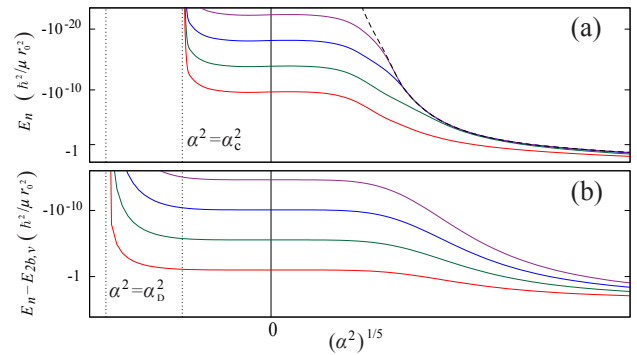}
  \caption{(Color online.)
    (a) The lowest four three-body energies calculated from Eq.~(\ref{HyperradEqn}) for $W_{00}$ from Eq.~(\ref{Eq:TwobodyReg})
    with a hard wall added at $R$=100$r_0$. 
    The dashed line denotes the lowest two-body threshold.
    (b) Schematic three-body binding energies assuming 
    there is a deeply-bound two-body state whose energy is independent of $\alpha$.
    \label{Fig:Spectrum}}
\end{figure}

We summarize the behavior of the 
three-body spectrum discussed so far in the $\alpha^2$$<$0 portion of Fig.~\ref{Fig:Spectrum}(a).
At $\alpha^2$=$ \alpha_c^2$, an infinite
number of three-body states appears with energies given by Eq.~(\ref{Eq:SubSpectrum}).  At $\alpha^2$=0, 
an infinite number of two-body states emerges --- only the lowest of which is indicated in Fig.~\ref{Fig:Spectrum}(a)
for simplicity.

Before considering the $\alpha^2$$>$0 behavior, let us revisit the zero-range two-body potential in Eq.~(\ref{Eq:Twobody}).
The theoretically attractive fact that $R$ can be exactly separated in the Schr\"odinger equation begs the question:
what are the $\alpha_\nu^2$ in Eq.~(\ref{Eq:Universal})?  It turns out that Eq.~(\ref{Eq:LongRange}) at a fixed $R$
provides the answer.  In the limit $r_0$$\rightarrow$0, the numerator --- proportional to $\alpha_0^2$ ---
diverges.  This conclusion is supported by taking the limit numerically for various regularization schemes.
While this divergence is a bit disappointing, it is also interesting in its own right as
a novel three-body ``fall-to-the-center'' problem that occurs even when there is no divergence in the two-body 
problem.

Returning to Fig.~\ref{Fig:Spectrum}(a), we see that
the three-body energies connect smoothly across $\alpha^2$=$0$.
The nature of the spectrum, however, changes.  For $\alpha^2$$>$0,
the three-body system can break up into a dimer and a free particle when $R$$\rightarrow$$\infty$. 
The leading term in $W_{\nu\nu}$ is then
given by the sum of the two-body interactions between the free particle and each of the particles 
in the dimer.
Thus, the asymptotic form for the $J^\Pi$$=$$0^+$ three-body potentials is
\begin{equation}
W_{\nu\nu}\rightarrow E_{vl}-\frac{\alpha_{\rm eff}^2+1/4}{2\mu R^2}\hbar^2,
\label{Eq:Super}
\end{equation} 
where
$E_{vl}$ is the energy of a two-body vibrational state $v$ with orbital angular momentum $l$ and
\begin{equation}
\alpha_{\rm eff}^2=\frac{8}{3}\alpha^2+\frac{5}{12}-l(l+1).
\label{Eq:Alpha3}
\end{equation}
Note that to have $J^\Pi$$=$$0^+$, $l$ must also be the orbital angular momentum of the free particle relative to the dimer.
We have verified Eq.~(\ref{Eq:Super}) numerically for the two-body potential in Eq.~(\ref{Eq:TwobodyReg}).

For $s$-wave dimers, Eq.~(\ref{Eq:Alpha3}) shows that for each two-body bound channel, 
the interaction in Eq.~(\ref{Eq:Super}) is always supercritical when $\alpha^2$$>$0. 
An infinite number of three-body bound states thus lie in the hyperspherical potential asymptoting to $E_{00}$
with energies relative to $E_{00}$ given by Eq.~(\ref{Eq:EfimovSpec}) after replacing $\alpha$ by $\alpha_{\rm eff}$.
But, there are also an infinite number of three-body resonances 
attached to each of the excited two-body thresholds $E_{v0}$ whose energies relative to
their respective threshold are also given approximately by Eq.~(\ref{Eq:EfimovSpec}).

For larger values of $\alpha^2$, it is possible to have an infinity of two-body dimers with
$l$$>$0.  In particular, when $l_0(l_0\!+\!1)$$<$$\alpha^2$$\leq$$(l_0\!+\!1)(l_0\!+\!2)$,
the effective two-body potential, including the centrifugal barrier, is supercritical for $l$$\le$$l_0$.
Equation~(\ref{Eq:Alpha3}) shows that so long as $\alpha^2$ is supercritical for a given $l$,
$\alpha_{\rm eff}^2$ will also be supercritical.  In other words, there will be infinite
series of states attached to every two-body threshold.  While the energies of all of these
series relative to $E_{vl}$ are given by Eq.~(\ref{Eq:EfimovSpec}), note that $\alpha_{\rm eff}$
in Eq.~(\ref{Eq:Alpha3}) does depend on $l$.

To this point, we have assumed that the short-range modification of the two-body potential only 
makes it shallower, thus preserving the property that no two-body states exist for $\alpha^2$$\le$0.
This assumption is convenient for establishing 
the characteristics of the system but is certainly
not necessary.  If we now allow short-range modifications that support deeply-bound two-body states,
then by the arguments above, it is clear that Eq.~(\ref{Eq:Super}) will also apply for $\alpha^2<0$. 
Because these deeply-bound two-body states are independent of $\alpha$, Eq.~(\ref{Eq:Alpha3}) implies a new critical
value for $\alpha^2$ determined from $\alpha_{\rm eff}^2$=0: $\alpha_D^2$=$3 l(l\!+\!1)/8\!-\!5/32$. 
The spectrum of these three-body states is given by Eq.~(\ref{Eq:EfimovSpec}), and they
will coexist with those states from the higher channel given by Eq.~(\ref{Eq:SubSpectrum}). 
We show the lowest few states of the three-body spectrum attached to this deeply bound two-body threshold in Fig.~\ref{Fig:Spectrum}(b).
There will be a similar spectrum for each deeply bound two-body state.

A natural question to ask is whether the effect we have described occurs for other systems such as three
identical bosons with $J$$>$0, different mass particles, or even combinations of identical fermions.
Since the latter seems to have some relevance to the work recently published in Ref.~\cite{Castin2010}, we
will focus on it.  

We have calculated $W_{\nu\nu}$ numerically for three indistinguishable, spin-polarized fermions
interacting via Eq.~(\ref{Eq:TwobodyReg}). In this case, we study the symmetry $J^\Pi$=$1^+$
since it gives the lowest potential in the non-interacting case~\cite{Esry2001}. 
The lowest two-body angular momentum satisfying the symmetry constraints is $l$=1, implying that $\alpha^2$ can be as large
as 2 without having a two-body bound state.  
Having examined $W_{00}$ for several
$\alpha^2$$<$2, we have empirically found that the asymptotic behavior of $W_{00}$ is
\begin{equation} 
 W_{00}(R)\rightarrow -\frac{\alpha_{\rm eff}^2+1/4}{2\mu R^2}\hbar^2 -\frac{\gamma}{2\mu \ln(R/r_0) R^2}\hbar^2.
\label{Eq:Fermion}
\end{equation}
For $\alpha^2$=$2$, we have obtained $\alpha_{\rm eff}^2$=$5.24$ and $\gamma$=$4.19$ by fitting Eq.~(\ref{Eq:Fermion}) to our numerical potential.
Therefore, our effect does occur: the three-fermion system has an infinite number of three-body bound states ---
in the absence of two-body bound states --- with a spectrum given approximately by Eq.~(\ref{Eq:EfimovSpec}).  Note that even
though our effect shares the same spectrum as the Efimov effect, it exists in a $1^+$ system of 
identical fermions where the Efimov effect does not~\cite{MacekFermionEfimovPRL}.

Based on these results, we believe that our effect might better explain
the four-body states recently found by Castin {\em et al.}~\cite{Castin2010}
than does the Efimov effect.  Castin {\em et al.} considered a system of
three heavy, spin-polarized, identical fermions $H$ plus a fourth lighter particle $L$,
assuming the only interactions were contact potentials between $H$ and $L$.  Upon
setting the $H$+$L$ $s$-wave scattering length $a_{HL}$ to infinity, they found a supercritical
$1/R^2$ interaction for 13.384$\le$$m_H/m_L$$\le$13.607 --- if the $H$s had $1^+$ symmetry.

To connect with our effect, we can approximately reduce their four-body problem to an effective $H$+$H$+$H$ problem 
by integrating out the $L$ motion via the Born-Oppenheimer approximation~\cite{Fonseca1979}.  
The resulting Born-Oppenheimer surface for most configurations of the three $H$s can be
well-approximated as a sum of $H$+$H$ pair potentials. 
Since $a_{HL}$=$\infty$,
Efimov's analysis tells us these pair potentials behave as $1/r^2$~\cite{Braaten2006}.
We thus have a three-body problem with attractive $1/r^2$ interactions ---
{\em i.e.} the focus of this Letter.  

Equation~(\ref{Eq:Fermion}) shows that, unlike the bosonic case, we can take the limit $r_0$$\rightarrow$0
for fermions to match the assumptions in Ref.~\cite{Castin2010}, leaving just the first term in 
Eq.~(\ref{Eq:Fermion}) in agreement with Eq.~(\ref{Eq:Universal}).
For $\alpha^2$=2, corresponding to $m_H/m_L$=13.607, we thus obtain for $\alpha_0^2$ from Eq.~(\ref{Eq:Universal}) the 
supercritical value of 5.24.
Reducing $\alpha^2$ until $\alpha_0^2$ is zero
gives $\alpha^2$$\approx$1.6 
--- the fermion equivalent of $\alpha_c^2$ found for bosons above.
Remarkably, we have thus found an infinity of three-body bound states when the 
effective two-body potential (including the centrifugal term) is repulsive! 
Using the zero-range model~\cite{NielsenReview,Braaten2006}, we can convert this to a mass
ratio of $m_H/m_L$=11.58, implying that an infinity of $1^+$ three-fermion bound states --- in
the absence of two-body states --- exists for 11.58$\le$$m_H/m_L$$\le$13.607.

It is clear that our range of mass ratios and values of $\alpha_{\rm eff}^2$ do not quantitatively
match those from Ref.~\cite{Castin2010}.  One of reasons for these quantitative differences
is likely the fact that $m_H/m_L$ is only roughly 10, reducing the reliability of
the Born-Oppenheimer approximation.
Another likely reason for the difference is the approximation of the
Born-Oppenheimer surface as purely a pair-wise sum of $1/r^2$ interactions when simple 
arguments suggest that the surface is less attractive for some configurations.
Nevertheless, we think our explanation of the infinity of states from Ref.~\cite{Castin2010} 
is compelling.  And, while they could not exist without the Efimov physics of the $H_2L$
system, they are not themselves Efimov states for all of the reasons we have discussed.

In summary, we have found a new class of three-body states: when three particles interact
via attractive $1/r^2$ potentials, they can
have an infinite number of three-body bound states even when there are no two-body bound states.
This effect occurs for both $0^+$ identical bosons and $1^+$ identical fermions --- and almost
certainly for many other cases as well.
For identical bosons, this effect produces a unique three-body spectrum and a new kind
of three-body fall-to-the-center problem.  For identical fermions, this effect sounds especially
surprising since it occurs when the effective two-body interaction is completely repulsive.
Moreover, this effect still occurs when there are two-body bound states.  While these states
do share some characteristics with Efimov states, their physical origin is quite distinct.
We use this distinctiveness to argue that the recently-found four-body ``Efimov'' states
are, in fact, ``our'' states --- arguments that seem likely to apply equally well to even more particles.

\begin{acknowledgments}
BDE and NG acknowledge support from the National Science Foundation Grant PHY0805278, and
YW acknowledges support from the National Science Foundation Grant PHY0970114.
\end{acknowledgments}

\end{document}